\documentclass[fleqn,10pt]{mystyle}

\usepackage[T1]{fontenc}

\usepackage{tcolorbox}
\usepackage{lmodern}
\usepackage[colorlinks=true,citecolor=blue,linkcolor=blue,urlcolor=blue]{hyperref}
\graphicspath{{/home/brehelin/HDR/}}

\newcommand{\parencite}{\citep}
\newcommand{\doc}{review}
\newcommand{\hdronly}[1]{}
\newcommand{\reviewonly}[1]{#1}
\newcommand{\ie}{\textit{i.e.}}
\newcommand{\eg}{\textit{e.g.}}
\newcommand{\vs}{\textit{vs.}}

\newlength{\mylength}
\newcommand{\coteacote}[3]{
  \begin{minipage}{#1\linewidth}#2\end{minipage}\setlength{\mylength}{\linewidth}\addtolength{\mylength}{-#1\linewidth}\begin{minipage}{\mylength}#3\end{minipage}}
\newcommand{\centre}[1]{\begin{center}#1\end{center}}
\newcommand{\dexter}{DExTER}
\newcommand{\Pf}{\textit{P. falciparum}}
\newcommand{\approach}[1]{\texttt{\underline{#1}}}
\tcbset{fontupper=\ttfamily,nobeforeafter,tcbox raise base,arc=0pt,outer arc=0pt,top=0pt,bottom=0pt,left=0mm,right=0mm,leftrule=0pt,rightrule=0pt,toprule=0pt,bottomrule=0.3mm,boxsep=0.5mm}

\title{Advancing regulatory genomics with machine learning}

\author[1]{Laurent Bréhélin}
\affil[1]{LIRMM, Univ Montpellier, CNRS, France}
\corrauthor[1]{Laurent Bréhélin}{brehelin@lirmm.fr}

\begin{abstract}
In recent years, several machine learning approaches have been
proposed to predict gene expression and epigenetic signals from the
DNA sequence alone. These models are often used to deduce, and, to
some extent, assess putative new biological insights about gene
regulation, and they have led to very interesting advances in
regulatory genomics. This article reviews a selection of these
methods, ranging from linear models to random forests, kernel methods,
and more advanced deep learning models. Specifically, we detail the
different techniques and strategies that can be used to extract new
gene-regulation hypotheses from these models. Furthermore, because
these putative insights need to be validated with wet-lab experiments,
we emphasize that it is important to have a measure of confidence
associated with the extracted hypotheses. We review the procedures
that have been proposed to measure this confidence for the different
types of machine learning models, and we discuss the fact that they do
not provide the same kind of information.
\end{abstract}

\begin{document}

\flushbottom
\maketitle
\thispagestyle{empty}

\section{Introduction}
Machine learning (ML) has been used for decades in genomics and
bioinformatics. Among many other examples, protein modeling with
Hidden Markov models dates back to 1992
\parencite{haussler_protein_1992}, and the book of P. Baldi and
S. Brünak \emph{Bioinformatics: the machine learning approach} was
published in 1998 \parencite{baldi_bioinformatics_1998}. In regulatory
genomics specifically, position weight matrices (PWMs), which are the
most common models for transcription factor (TF) binding sites,
appeared in the late 80's \parencite{schneider_information_1986,stormo_consensus_1990}. In the following years, many
algorithms have been proposed in the literature to estimate the
parameters of a PWM from sequence examples
\parencite{stormo_consensus_1990, bailey_fitting_1994,
  ruan_comparison_2018,bailey_streme_2021}. Today, PWMs of hundreds of
TFs are available in databases like JASPAR
\parencite{fornes_jaspar_2020} and HOCOMOCO
\parencite{kulakovskiy_hocomoco:_2017}, and can be used to compute
binding affinities and to identify potential binding sites in genomes.

In recent years, several ML approaches have been proposed to go beyond
single TF binding sites, by modeling entire regulatory sequences
spanning hundreds or even thousands base pairs. These models range
from linear models \parencite{vandel_probing_2019} to random forests
\parencite{whitaker_predicting_2015}, kernel methods
\parencite{lee_discriminative_2011,ghandi_enhanced_2014},
convolutional neural networks
\parencite{alipanahi_predicting_2015,zhou_predicting_2015}, and more
advanced deep learning approaches
\parencite{quang_danq:_2016,vaswani_attention_2017}. Notably, deep
learning approaches adapted from methods initially developed for image
and natural language processing have attracted considerable attention
in the field. These studies take place in a supervised framework,
where the goal is to train a model able to predict a signal measuring
gene expression (RNA-seq, CAGE, \dots), TF binding or histone marks
(ChIP-seq, ATAC-seq, \dots) on the basis of the DNA sequence
only. Despite the supervised framework, in a large number of studies
these models are not really used as predictors. Instead, the goal is
to use the model to deduce, and to some degree assess, new biological
knowledge and hypotheses about gene regulation; hypotheses that have
then to be experimentally validated.  Along with the availability of a
huge quantity and diversity of genomic, transcriptomic and epigenetic
data, these approaches have allowed very interesting progresses in
regulatory genomics.  Maybe one of the most striking results is the
fact that gene expression can be predicted with often high accuracy
from the sequence only, which means that a large part of the
instructions for the control of gene expression likely lie at the
level of the DNA. This is in contradiction with a common belief that
gene expression first depends on chromatin marks that are not
necessarily controlled by the DNA sequence
\parencite{zeitlinger_seven_2020}. Actually, several of the early
works in the field have shown that epigenetic marks can be predicted
from the sequence alone, often with very good accuracy
\parencite{ghandi_enhanced_2014,whitaker_predicting_2015,zhou_predicting_2015}.

Beyond this very general result, other biological hypotheses can be
deduced from ML models. As we will see in this \doc, different
strategies/methods can be used to deduce and assess interesting
biological hypotheses with these models. Certain general knowledge,
such as the cell specificity of a regulatory mechanism, can be quite
easily tested by learning different models for different conditions
and comparing their accuracy. However, for more specific knowledge,
such as the DNA motifs involved in a specific regulation, we have to
directly analyse the learned model, and this knowledge-extraction
process is highly dependent on the model type. In addition, because
this putative knowledge must be validated with wet-lab experiments, it
is important to have a measure of importance associated with each of
the extracted hypotheses. However, we will also see that the different
models do not provide the same kind of importance
measure. Specifically, for some models these measures can be
considered as objective, in the sense that they rely directly on the
error of the model, estimated from the data. For other approaches,
however, these estimates are more subjective, as they are computed
solely on the basis of the signal predicted by the model, with no link
to the associated error.

\reviewonly{\bigskip There are several recent reviews devoted to deep learning
  approaches for modeling genomic sequences
  \parencite{eraslan_deep_2019,koo_deep_2020,talukder_interpretation_2020,routhier_genomics_2022}. In
  addition, several other papers address the problems of model
  interpretation in machine learning---see for example
  \parencite{montavon_methods_2018,lipton_mythos_2018,gilpin_explaining_2019,ghorbani_interpretation_2019}
  This review differs from these papers on several points: i) it is
  specifically dedicated to the modeling of genomic sequences involved
  in the regulation of gene expression; ii) it presents a wide variety
  of models, without being restricted to deep learning approaches;
  iii) it makes an inventory of methods and practices that can be used
  to extract different levels of biological knowledge from these
  models. This review is intended to biologists and computational
  biologists who are curious to know how machine learning can be used
  to deduce new biological hypothesis about gene regulation. Special
  attention has been paid to explaining the mathematical concepts
  simply and intuitively but also in a sufficiently detailed way, so
  that readers without knowledge of machine-learning should be able to
  fully understand the advantages and limitations of the different
  approaches. This paper does not intend to provide an exhaustive list
  of methods and approaches that use machine learning to study
  regulatory sequences. Rather, the aim is to develop and provide the
  reader with the technical concepts necessary to understand this
  literature. Most of the works presented below have been selected for
  their pioneering aspect, in the sense that they are, to our
  knowledge, the first to introduce a specific type of machine
  learning model or to apply a specific method to evaluate new
  hypotheses or to interpret a given machine learning model.}

This \doc\ is organized as follows. We first present a selection of
models for predicting either epigenetic marks or expression
signals. The following sections discus how these models are used to
infer new biological hypothesis, and give some examples of knowledge
that were gained from these studies. Specifically, we give a brief
literature tour on the identification and prioritization of nucleotide
variants. Next, we describe simple approaches that are used to assess
very general hypotheses with these models.  Finally, the last section
is devoted to the extraction of more complex DNA features, and to the
measures of importance that are associated with these features.

%% \reviewonly{\section{Survey methodology} This paper does not intend to
%%   provide an exhaustive list of methods and approaches that use
%%   machine learning to study regulatory sequences. Rather, the aim is
%%   to develop and provide the reader with the technical concepts
%%   necessary to understand this literature. Most of the works presented
%%   below have been selected for their pioneering aspect, in the sense
%%   that they are, to our knowledge, the first to introduce a specific
%%   type of machine learning model or to apply a specific method to
%%   evaluate new hypotheses or to interpret a given machine learning
%%   model. Some other papers were selected because they provide
%%   interesting illustrations of the kinds of analyses that are possible
%%   with these approaches.}

\section{Machine learning for regulatory genomics}\label{s:models}

Given a genome-wide experiment (RNA-seq, ChIP-seq, ATAC-seq, \dots)
monitoring a specific signal (gene expression, TF binding, histone
mark, \dots) in a specific condition (cell type, time point,
treatment, \dots), the aim is to learn a model able to predict this
signal based on the DNA sequence alone. We have a set of
sequences $X=\{x^1,\dots,x^N\}$, each sequence $x^i$ being associated
with a signal $y^i$. For classification problems (typically when
predicting TF binding or histone mark) $y^i$ is limited to two values
(\eg\ -1 / +1) indicating whether the $i$-th sequence is or is
not bound by the studied factor in the ChIP-seq experiment. For
regression problems (typically when predicting a gene expression
signal) $y^i$ is a continuous value that measures the expression
associated with $i$-th sequence (in this case, sequences may be for
example gene promoters or enhancer sequences). Therefore, the goal is
to learn a prediction function $f(x)$ that predicts the signal
associated with sequence $x$. An important remark is that the function
$f$ is associated with a specific experiment and therefore with specific
conditions: if the conditions change (for example if the cell is
treated with a new drug) the function $f$ may no longer be a good fit
of the cell and another predictor should be learned from data
monitoring these new conditions.

Note that even when the purpose of the study is not to make
predictions \textit{per se}, the prediction framework keeps several
interests: first, when estimating the accuracy of the model we also
assess the amount of information that has been captured by the
model---and, maybe more importantly, what remains to be
captured. Second, in comparison to simple correlation analyses that
can also identify some links between sequence features and the target
signal, predictive models enable the combination of several DNA
variables and hence reveal cellular mechanisms that cannot be studied
with simple correlation analyses.

In the following, we distinguish the methods that predict an
expression signal from those that predict epigenetic marks.

\subsection{Predicting epigenetic marks}

\subsubsection{Support vector machines}
One of the first studies dedicated to the prediction of ChIP-seq
signal based on long regulatory sequences was the kmer-SVM approach
\parencite{lee_discriminative_2011}, latter upgraded in gkm-SVM
\parencite{ghandi_enhanced_2014}. Support vector machines (SVMs) are
among the most successful methods of ML
\parencite{cortes_support-vector_1995}. They work by searching for an
hyperplane that best separates the training examples according to
their classes. However, rather than searching for an hyperplane in the
original space of the data, SVMs work in a usually much higher
dimensional space (with possibly infinite dimension). As a result,
while the separating hyperplane is by definition linear in the
enlarged space, it can define non linear boundaries in the original
space. The beauty of SVMs is that we do not need to define the
enlarged space. The only thing we have to define is a \emph{kernel}
function that measure the similarity of any pair of examples. The
"trick" is that the position of a new example relative to the
separating hyperplane can be computed by a linear combination of the
similarity of this new example with all training examples, \ie{}
\begin{equation}
  f(x) = b + \sum_{i=1}^N \alpha_i \, y^i\, K(x,x^i),\label{eq:SVM}
\end{equation}
with \(b\) and \(\alpha_i\) the parameters of the SVM estimated by the
learning algorithm, \(y^i\) (-1/+1) the class of the $i$-th sample of the
training set, and \(K(x,x^i)\) the result of the kernel function
between sample \(x\) and the $i$-th example of the training set. The
sign of \(f(x)\) gives us the position of \(x\) relative to the
hyperplane (up or down) and hence the predicted class of the
sample. Moreover, only some training examples are associated with a
\(\alpha_i\) different from 0 (the so-called \emph{support vectors}),
so in practice Expression (\ref{eq:SVM}) is computed without the
need to compute the kernel function for all training examples.

Thus, the accuracy of the approach all depends on the chosen kernel
function \(K(x,x')\) which must be meaningful for the problem at
hand. The authors of \approach{kmer-SVM}
\parencite{lee_discriminative_2011} proposed a kernel function based
on the similarity of the $k$-mers present in the sequences. More
formally, each sequence $x$ is encoded by a vector that reports the
number of occurrences of each of the $4^k$ $k$-mers in $x$ (for a
given $k$-mer size $k$). Sequence pairs that have the same relative
frequency for each $k$-mer have $K(x,x')=1$, while sequences that have
not any $k$-mer in common have $K(x,x')=0$. The method has been
improved a few years later by the \approach{gkm-SVM} approach that uses
gapped $k$-mers (\ie\ $k$-mers with a certain number of
non-informative positions for which any nucleotide is possible)
\parencite{ghandi_enhanced_2014}, but the principle remains the same:
training sequences that most resemble the new sequence $x$ in terms of
(gapped) $k$-mers composition have more weight in
Expression~(\ref{eq:SVM}), and drive the predicted class toward their
own class. The gkm-SVM approach has been applied to 467 human ChIP-seq
experiments from the ENCODE project
\parencite{encode_project_consortium_encode_2004}. For each ChIP-seq
experiment, sequences associated with a ChIP-seq peak were extracted
and used as positive sequences (authors report an average sequence
length of around 300bps), while the same number of random genomic
sequences were used as negative sequences. Then, an SVM (one for each
ChIP-seq experiment) was learned to discriminate between these two
sets, and the accuracy of each of these 467 SVMs was estimated by the
Area Under the Receiver Operating Curve (AUROC). The approach showed
very good results, with AUROC above 80\% for most
datasets. As for any ML method, to avoid any optimistic bias the
accuracy estimate must be computed on left-out sequences,
\ie\ positive and negative sequences that have not been used to train
the model.  The number of parameters of the SVM ($b$ and $\alpha_i$ in
Expression~(\ref{eq:SVM})) depends on the number of training
examples. It is thus directly linked to the number of ChIP-seq peaks
of the experiment, which usually varies from a few to several
thousand.

\subsubsection{Random forests}
\begin{figure}
  %\footnotesize
  \tiny
  \begin{center}
    \begin{tabular}{l|c|cccccc}
      \hline
      & y     & \textbf{pwm\#1} & \textbf{pwm\#2} & \textbf{pwm\#3} & \textbf{pwm\#4} & \textbf{\dots} & \textbf{pwm\#50} \\ 
      \hline
      \textbf{seq. 1}    & 1     & 15               & 24               & 17               & 24               & \dots          & 26               \\
      \textbf{seq. 2}    & 0     & 9                & 21               & 20               & 19               & \dots          & 12               \\
      \textbf{seq. 3}    & 0     & 17               & 19               & 16               & 18               & \dots          & 24               \\
      \textbf{\dots}     & \dots & \dots            & \dots            & \dots            & \dots            & \dots                             \\
      \textbf{seq. 1000} & 1     & 18               & 18               & 17               & 20               & \dots          & 15               \\
      \hline
    \end{tabular}\small

    \includegraphics[width=.5\linewidth]{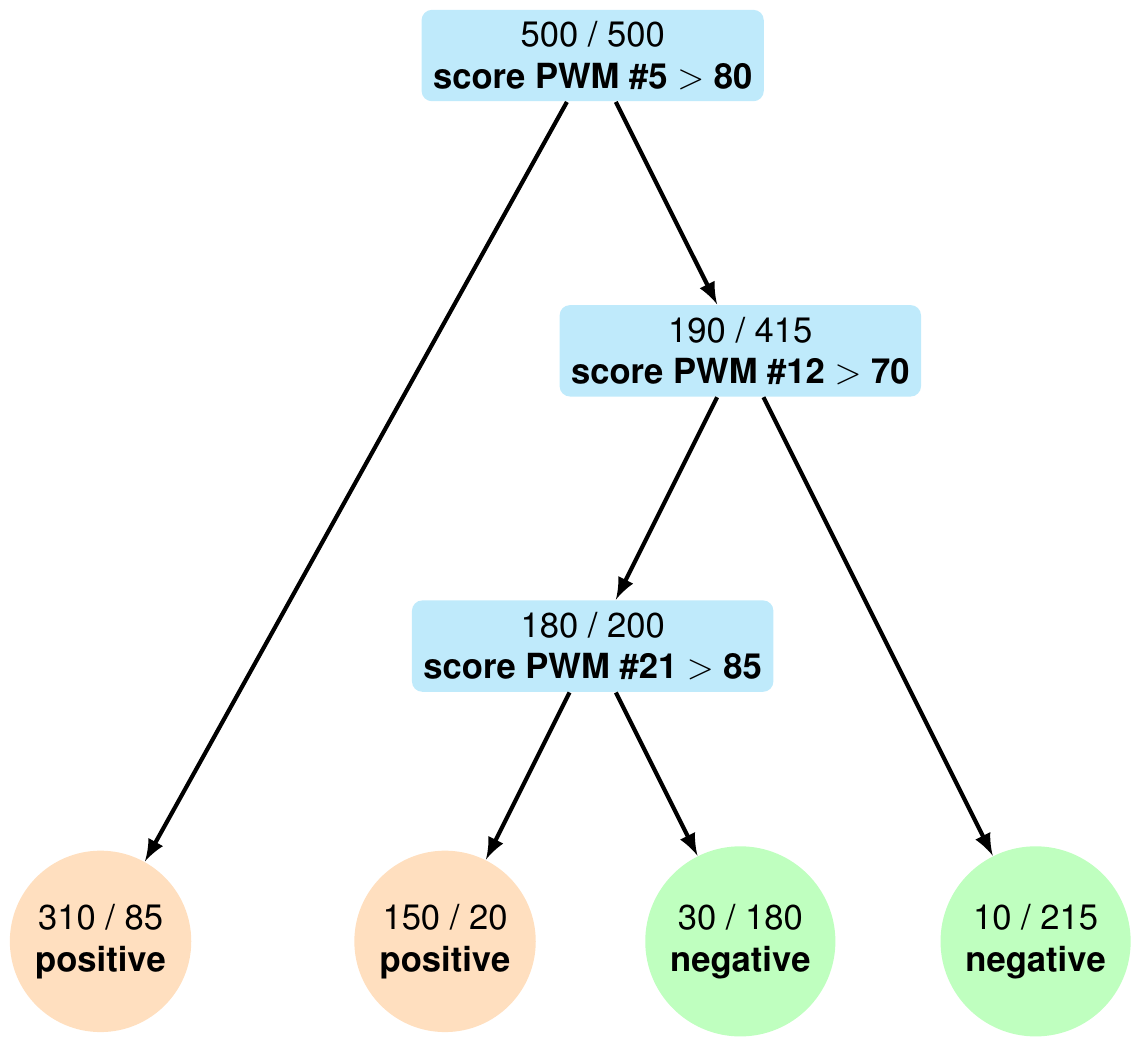}
  \end{center}
  \caption{\textbf{A toy decision-tree inspired from the Epigram
      method \parencite{whitaker_predicting_2015}.} The learning set involves 1000 sequences
     (500 positive $+$ 500 negative, see top
    table). Each sequence is described by the score of 50 PWMs. The
    decision tree learned from these sequences encodes a set of rules
    that classify a sequence as positive if the score of PWM$\#5$ is
    $>$ 80, or if the score of PWM$\#12$ and $\#21$ are $>$ 70 and
    85, respectively. Numbers at the top of each node provide the repartition
    of training sequences in the two classes.}\label{fig:tree}
\end{figure}

Another method that has been proposed for the prediction of ChIP-seq
signal is random forests (RFs)
\parencite{whitaker_predicting_2015}. RFs
\parencite{breiman_random_2001} got considerable success and attention
in the ML literature in the last twenty years. RFs are an extension of
\emph{decision trees}, which are certainly among the oldest approach in
ML \parencite{morgan_problems_1963}. A decision tree encodes a set of
tests that can be done on the features of a given example to predict
its class (see Figure~\ref{fig:tree}). As the name suggests, these
tests are organized in a tree structure that provides the order in
which the tests are done (from the root to the leaves). Each node
corresponds to a binary test on a specific feature of the samples (for
example, ``is feature \#3 $>$ 0.66?''), and the two edges that follow
the node correspond to the two possible outcomes of the test. Thus,
each path from the root to a leaf corresponds to a series of tests
that lead to the prediction of the class associated with that leaf
(which is the majority class of the training examples that belong to
this leaf). Learning a decision tree involves deciding the tests
associated with each node (typically which feature and which threshold
on this feature?), and several algorithms have been proposed to build
trees with minimum prediction error. Decision trees are accurate on
certain problems and have the great advantage of being simple to
understand and interpret. However, they are known to be quite unstable
(a slight change in the training data can lead to very different
trees) and they can be relatively inaccurate on some problems
\parencite{breiman_random_2001}. Random forests have been proposed to
overcome these issues, at the price however of an indubitable loss of
readability. A RF is a combination of many (sometimes hundreds or even
thousands) decision trees. All these trees are independently learned
on the same problem, with a dedicated algorithm that integrates a
stochastic procedure to ensure that all trees are different.  Once
learned, the RF can be used to predict the class of a new sample with
a very simple procedure: each tree is used to predict the class, and
the most popular class is proposed by the forest. This voting scheme
removes the instability problem, and, if the learned trees are
sufficiently independent, the accuracy of the forest can be much
higher than that of single trees \parencite{breiman_random_2001}. The
independence is ensured by two stochastic components of the learning
algorithm: 1) each tree is learned on a different bootstrap sample of
the original data, which means that the learning set is obtained by
randomly sampling with replacement the original learning set (in this
way, some samples may appear several times in the bootstrap sample,
while others are absent); 2) at each node, only a randomly chosen
subset of features are considered to determine the exact test
associated with the node.

\bigskip
The \hypertarget{epigram}{\approach{Epigram}} approach \parencite{whitaker_predicting_2015}
uses RFs to predict the presence of six histone marks in different
human cell types from DNA sequence. A RF is learned for each mark
based on the peaks identified by dedicated ChIP-seq
experiments. Sequences associated with the peaks were used as positive
sequences (sequence length is around 1000bp). Negative sequences were
genomic regions not covered by a peak. Before learning a RF, a
\textit{de novo} motif finding algorithm was used to identify motifs
that are more present in one class than in the other. Then, each
sequence was described by the score obtained by the different motifs,
and the RF learning algorithm was run on these data. As for SVM, the
accuracy of the approach was measured by the AUROC and showed
very good results above 80\% for most histone marks and cell types.
Determining the number of parameters of a RF is tricky, as it depends
on the number of nodes of the trees. If we assume that this number is
of the order of the number of features considered, for Epigram it
would be around 300. The number of trees of the forest is 1000, so we
can assume that the number of parameters of each RF is of the order of
300K.

\subsubsection{Convolutional neural networks}
\begin{figure}
  \includegraphics[width=.5\linewidth]{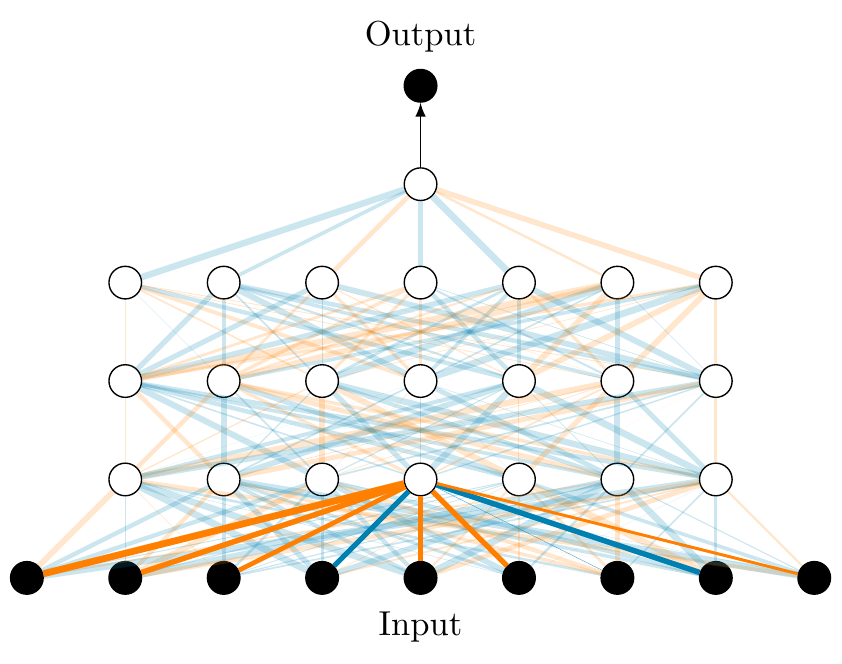}
  \includegraphics[width=.5\linewidth]{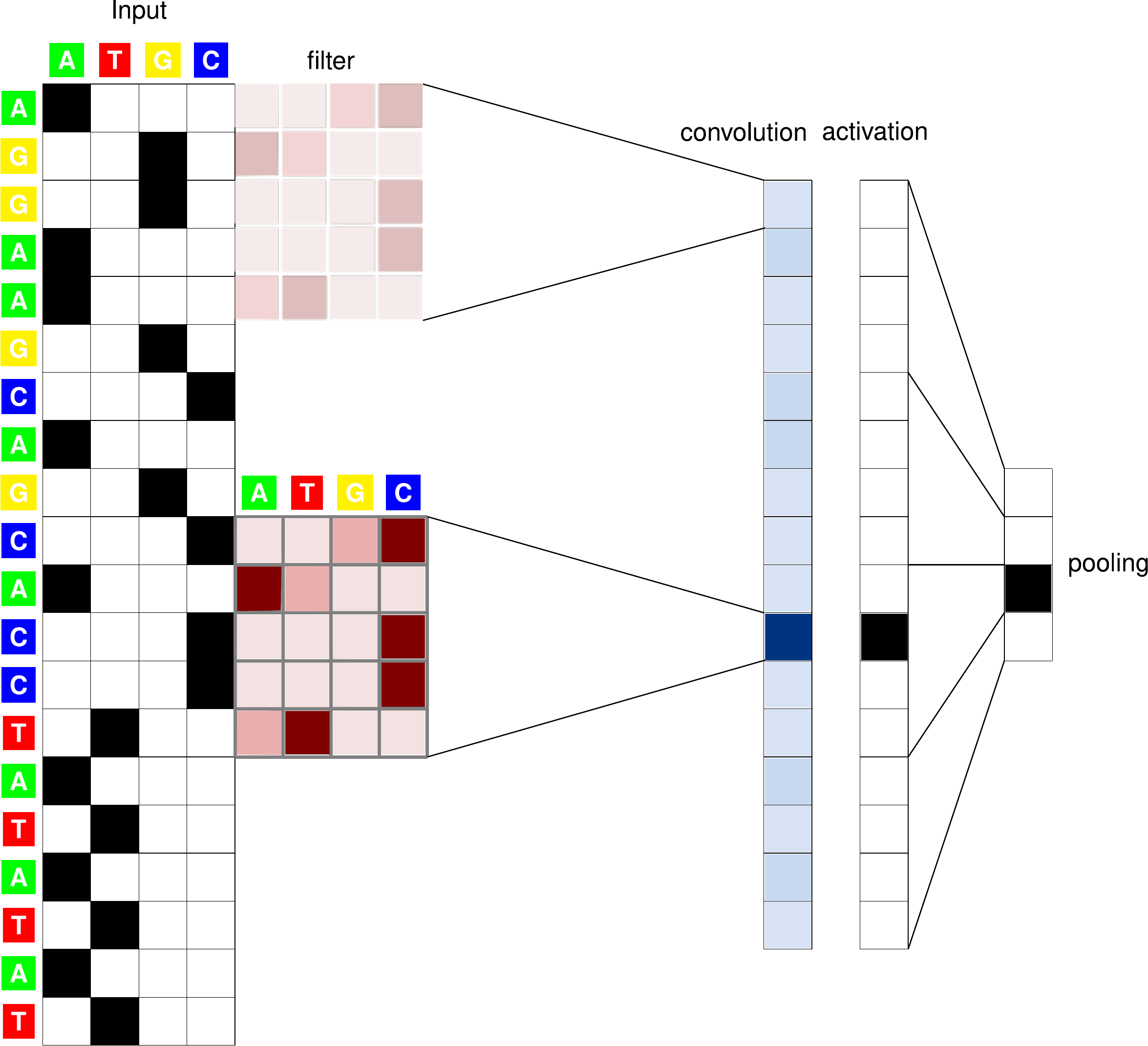}
  \coteacote{.5}{\centre{(A)}}{\centre{(B)}}
  \caption{\textbf{Neural networks.}\textbf{(A)} An example of
    feedforward network with 4 layers. The first layer is connected to
    9 input values (in black). The last layer involves a single neuron
    that outputs a value between 0 and 1. The width and color of the
    lines represent the weights and signs associated with the inputs
    to each neuron. \textbf{(B)} An extract of a convolutional neural
    network that scans DNA sequences. The input sequence is encoded
    as a one-hot matrix. A filter of length 5 scans the inputs and
    computes a convolution at each position. A non-linear activation
    function then filters these values by removing all values below a
    given threshold. Finally, the results of these operations are
    pooled by groups of 4, reducing the output of the convolution to
    only 4 values. On the figure, only one filter is represented,
    although each convolution layer usually involves dozens or hundreds
    filters with their own activation and pooling operations. The
    results of all filters are then combined in the following layers
    of the network (not represented here).}\label{fig:NN}
\end{figure}

After SVMs and RFs, several authors proposed to predict TF binding and
histone marks with deep neural networks
\parencite{alipanahi_predicting_2015,zhou_predicting_2015}. The
simplest form of neural networks is the feedforward neural networks
(see Figure~\ref{fig:NN}.A). These networks are defined by a
(potentially very large) set of neurons, inter-connected and organized
in layers. The first layer is connected to the input values of the
network, while the last layer encodes its output. In the classical
feedforward network, each neuron takes in input the output of the
neurons of the previous layer. It then computes a weighted sum of
these values using its own set of weights, applies a simple activation
function that realizes a non-linear transformation of this sum, and
dispatches the computed value to the neurons of the following
layer. When new values are applied to the input layer of the network,
the outputs of all neurons are computed iteratively layer by layer
until reaching the last one. For classification problems, the last
layer is usually composed of a single neuron that produces a value in
between 0 and 1 representing the probability that the example provided
in input belongs to the positive class. The architecture of the
network (the number of layers, neurons, and all the connections
between them) as well as the set of weights associated with each
neuron and the form of the activation functions define the parameters
of the network. As for RFs and SVMs, specialized learning algorithms
can be used to train a neural network by minimizing its prediction
error on learning examples. Deep learning has shown considerable
success in image recognition and many other domains, including biology
and regulatory genomics. It has also led to very interesting
developments in ML theory, by showing that models with a number of
parameters largely exceeding the number of learning examples can be
trained in certain conditions without being affected by the so-called
\emph{over-fitting problem}
\parencite{nakkiran_deep_2019,belkin_two_2020}.

In regulatory genomics, most neural networks are convolutional neural
networks (CNNs). Contrary to simple feedforward networks, CNNs also
embed specific neurons called \emph{filters}, that realize
convolutional operations (see Figure~\ref{fig:NN}.B). This operation is very similar to the
weighted sum of classical neurons, except that the sum does not
involve all outputs of the previous layer but only a small group of
adjacent outputs. Moreover, this computation is repeated at every
position of the previous layer. Hence, one filter produces a number of
outputs that is roughly equal to the size of the previous layer
(modulo the few positions on the edges of the layer that correspond to
the filter size). As for classical neurons, the parameters of the
filters, \ie\ their sizes, forms and weights, must be inferred by the
learning algorithm. However, because the same filter is applied to
every position of the previous layer, convolution filters effectively
constitute a way to reduce the number of parameters of the model when
the same local operations is applied to different regions of a
layer. CNNs typically possess one or several convolutional
layers. Each layer involves several convolutional filters that are
applied in parallel. As for classical neurons, all outputs of a
convolutional filter are run through a non-linear activation
function. Contrary to classical neurons, however, this step is often
followed by a pooling function that summarizes (pools) several
adjacent outputs. One of the most common pooling functions is max
pooling, which simply takes in input a number $N$ of adjacent outputs
of the filter, and returns a single value corresponding to the maximum of
these values.

The big interest of CNNs is that, contrary to other ML approaches, the
network takes raw data in input and  directly learns the most
interesting features for discriminating the two classes by way of the
filters of the convolutional layers. In regulatory genomics, the raw
data is the DNA sequence. As CNNs only work on numbers, the sequence
is \emph{one hot encoded} by replacing each nucleotide with a boolean
vector of size 4 (for example A, T, G and C are encoded as 1000, 0100,
0010 and 0001, respectively). Hence, a DNA sequence of length $M$ is
encoded as a $4 \times M$ boolean matrix (see
Figure~\ref{fig:NN}.B). The CNNs used for regulatory genomics have
usually one or more convolutional layers, followed by several fully
connected feedforward layers. The filters associated with the first
layer are thus directly applied to the sequence and can be viewed as
models of DNA motifs. Actually, the weighted sum of a filter defines
exactly the same operation as the function used to compute the score
of a sequence for a given PWM. Hence, the filters of the first layer
are nothing more than a set of PWMs that are applied at each position
of the sequence. All these scores are then combined in the following
layers, which allows the network to represent potentially any motif
combination. Concerning the complexity of the model, the number of
parameters of a CNN depends on the input size, on the size and number
of filters, and, more importantly, on the shrinkage proportion of the
pooling operations and  the number of neurons in the fully connected
layers.

\bigskip

One of the first CNN approaches proposed for regulatory genomics was
\hypertarget{deepbind}{\approach{DeepBind}} \parencite{alipanahi_predicting_2015}. The authors
designed a CNN with one convolutional layer, followed by a pooling
layer and a feedforward network that combines the scores of the
different filters. The network was trained on many  datasets
measuring the binding of various DNA and RNA-binding proteins (PBM,
SELEX, ChIP/CLIP). Notably, it was trained on 506 ChIP-seq data from
the ENCODE project \parencite{dunham_integrated_2012}. Positive sequences
were 101bp sequences associated with a ChIP-seq peak, while the
negative sequences were obtained by randomly shuffling the
dinucleotide of the positive sequences. As for SVMs and RFs, the
accuracy of the CNNs was estimated by the AUROC. According to the
architecture provided in the paper (filter size: 24; \#filters: 16;
total pooling; 1 layer of 32 fully connected neurons), we can estimate
that the number of parameters of a DeepBind model is around $2K$.

\bigskip

The same year, Zhou \& Troyanskaya proposed another CNN approach named
\hypertarget{deepsea}{\approach{DeepSEA}} \parencite{zhou_predicting_2015}. DeepSEA has three
convolutional/pooling layers followed by a fully connected layer. It
was applied to 690 TF binding profiles, 125 DHS profiles and 104
histone-mark profiles from ENCODE and the Roadmap Epigenomic projects
\parencite{dunham_integrated_2012,kundaje_integrative_2015}. For each
predictor, the positive sequences were the sequence associated with
the mark, while the negative sequences were randomly selected among
the sequences that are not associated with the mark but that are
associated with another mark in another data. CNN accuracies were
estimated by the AUROCs. DeepSEA takes input sequences of 1kb length,
much longer than DeepBind (100bp) but very similar to the size of the
sequences used by Epigram (see above). While a direct comparison of
DeepSEA and Epigram has not been explicitly done, both approaches
report an AUROC of $\sim 85\%$ on average for predicting histone marks
on comparable positive and negative sequences. With its 3
convolutional layers with several hundred filters, the low shrinkage
proportion of its pooling layers and its fully connected layer of 925
neurons, the number of parameters of the DeepSEA model can be
estimated to $\sim$60M. Note however that the same model predicts one
value for each signal (multi-task learning).

\bigskip

Following DeepSEA, Quang \& Xie proposed \hypertarget{danq}{\approach{DanQ}}
\parencite{quang_danq:_2016}. DanQ has a single layer of convolutions that
identify the motifs present in the input sequence, but this
convolution is followed by a layer of recurrent neural network known
as long short term memory (LSTM). As for convolutional filters, in a
recurrent network the same function is applied to every
input. However, contrary to CNN, this function also has a memory of
its previous outputs. In this way, the computed output value depends
both on the current input and on the previous outputs. In DanQ the
LSTMs take in input the result of the convolution filters at every
position of the sequence. Hence, the output of the LSTM depends both
on the value of the filters at the current position and on the output
of the LSTM at the previous positions. Moreover, DanQ implements two
LSTMs: one that reads the sequence forward, and one that reads the
sequence backward. The LSTM layer is then followed by a fully
connected layer. DanQ was applied exactly to the same data as DeepSea.
Its accuracy, measured by AUROC and by the area under the
precision-recall curve (AUPRC), showed a slight but constant
improvement over DeepSea. The DanQ model has only one convolutional
layer and the total number of parameters can be estimated at about
1M. As DeepSea, DanQ implemented a multi-task learning so the same
model predicts one value for each signal.

\bigskip

Other CNN approaches have been proposed in the following years for
predicting TFBS and chromatin marks. For example, the authors of
\hypertarget{bpnet}{\approach{BPNet}} slightly change the goal of the approach
\parencite{avsec_base-resolution_2021}. Rather than addressing a simple
classification problem where the aim is to predict whether or not a sequence
has a specific mark/TF, the goal of BPNet is to predict the
ChIP-seq profile of the sequence, \ie\ the expected number of reads at
each base pair. According to the authors, training the model on the
profile enables them to capture subtle regulatory features that are
not captured when training on binary signal, a behavior also reported
in  \parencite{toneyan_evaluating_2022}.

\subsubsection{Logistic models}
Following the former CNN approaches, our group proposed to predict TF
binding with an approach based on a logistic model named TFcoop
\parencite{vandel_probing_2019}\hdronly{ (see Appendix B)}. Logistic
models are one of the oldest and most used approaches for binary
classification problems. For a given example $x=(x_1,\dots,x_M)$
described by $M$ variables, a logistic model expresses the probability
that $x$ belongs to the first class with a linear expression
\begin{equation}
 P(1|x) = S\left(a + \sum_{i=1}^M b_i\cdot x_i\right),\label{eq:logistic}
\end{equation}
where $P(1|x)$ is the probability that example $x$ belongs to the
first class, $S$ is the sigmoid function, and $a$ and $b_i$ are the
regression coefficients which constitute the parameters of the model.
While being quite simple, logistic and generalized linear models have
gain considerable interest in domains with high-throughput data (like
genomics), thanks to the development of modern regularization methods
\parencite{friedman_regularization_2010}. For problems with high amount of
data, and in situations where the number of variables (\ie\ $M$ in
Expression~(\ref{eq:logistic})) may be larger than the number of
examples, these models can be trained very quickly and without being much
affected by over-fitting. The LASSO penalty especially
\parencite{tibshirani_regression_1994}, is a powerful regularization
technique that allows one to train a model and select the most
important features at the same time---\ie\ many regression
coefficients of Expression~(\ref{eq:logistic}) ($a$ and $b_i$) are set
to zero by the learning algorithm---, which is of obvious interest
when we are also interested to understand how a predictor works \parencite{zhao_model_2006}.

\bigskip
As for SVMs and RFs (and contrary to CNNs) the accuracy of the
approach essentially depends on our ability to provide to the learning
algorithm a set of variables that contains meaningful information for
the discrimination problem. In \hypertarget{tfcoop}{\approach{TFcoop}}
\parencite{vandel_probing_2019}, each sequence was described by the
affinity scores of all PWMs of the vertebrate JASPAR database
\parencite{fornes_jaspar_2020}, and by the relative frequency of every
dinucleotide. Affinity scores and dinucleotide frequencies were
computed on the entire sequences. The data thus resembles that used in
Epigram \parencite{whitaker_predicting_2015}, except that in this
latter the PWMs were trained \textit{de novo} and the dinucleotides
were not used as predictive variables. TFcoop was applied on 409
ChIP-seq experiments from ENCODE. The positive sequences were either
the promoters (+/- 500bp around the TSS) or the enhancers (+/- 500bp
around the peaks defined by the FANTOM5 project
\parencite{andersson_atlas_2014}) with a peak in the studied ChIP-seq
experiment, while the negative sequences were the promoters or
enhancers without a peak. The goal of TFcoop was to study the rules
governing TF combinations in promoters and in enhancers, as well as
for different gene categories. The accuracy of TFcoop was assessed by
the AUROC. Our experiments on the same sequence sets showed that its
accuracy was close to that of DeepSea. The complexity of the model was
however very different, as the number of parameters of TFcoop was
roughly equal to the number of PWMs of the JASPAR library, which was
$\sim$700.

Recently, we proposed another logistic+LASSO model named
\hypertarget{tfscope}{\approach{TFscope}} whose aim is to identify the key regulatory elements
that differentiate two ChIP-seq experiments
\parencite{romero_systematic_2022}\hdronly{ (see Appendix E)}. TFscope
is used to analyze the binding differences of one TF in two 
conditions (two cell types or two treatments), or of two paralogous
TFs with similar PWMs. Contrary to the above approaches, in TFscope
the positive and negative sequences are the sequences associated with
a peak in the first and second ChIP-seq experiments, respectively. As
in TFcoop, the logistic model integrates affinity scores of JASPAR
PWMs as well as the nucleotide environment of the sequence. However,
these variables are not simply computed on the entire sequences but on
different regions of interest identified by TFscope. Moreover, the
logistic model also integrates the score of a discriminative PWM which
is directly learned from the sequences, in order to identify the
subtle differences that may exist between the binding sites of the
target TF in the two experiments. With this model, we were able to
discriminate the binding sites of the same TF in different cell types
with often good accuracy (AUROC > 80\%), and to identify the
genomic features the were associated with these binding differences
(see last section). The number of parameters of TFscope is slightly
higher than that of TFcoop because there are more variables describing
the nucleotidic environment, but practically the total number of
parameters is always below 1000.

\subsection{Predicting expression signal}
Following the prediction of TF binding and chromatin marks, several
approaches have been proposed to directly predict gene expression from
the sequence. While for the former problem the sequence length rarely
exceeds 1000 bp, for expression signals there is a clear tendency to
propose new approaches that can handle increasing sequence lengths to
capture long-range interactions with distant regulatory sequences
(enhancers/silencers).

\subsubsection{Linear models}
In 2018, our group proposed a linear regression model
\parencite{bessiere_probing_2018}\hdronly{ (see Appendix A)}, for predicting the RNA-seq signal
associated with a gene from the nucleotide and dinucleotide
frequencies computed on different DNA regions of the gene: its core
promoter (+/- 500bp around TSS), distal upstream (-2000bp before TSS)
and downstream (2000bp after TSS) promoter regions, 5' and 3' UTRs,
exons, and introns. The aim was to study the links between gene
expression and the nucleotidic/dinucleotidic composition of specific
parts of the genes. With this approach, each gene is described by a
vector of $\sim$150 variables, and a linear model with LASSO penalty
is trained to predict the expression signal (hence, each model
involves $\sim150$  parameters). This approach was applied to 241 human
RNA-seq datasets from the TCGA database
\parencite{mclendon_comprehensive_2008}, and each model was assessed by
computing the correlation between the measured and predicted
expression signal. This approach showed modest accuracy (50-60\% correlation)
but still very surprising  given the simplicity of the
features used for prediction, and questioned  our understanding
of the role that simple nucleotidic and dinucleotidic enrichments may have
on the control of gene expression.

\bigskip
In 2021, we proposed a second method named \hypertarget{dexter}{\approach{\dexter}} that aimed
to predict gene expression from the presence of low complexity
sequences in promoters
\parencite{menichelli_identification_2021}\hdronly{ (see Appendix
  C)}. \dexter\ takes 4Kb sequences centered on the TSS (or gene
start) and identifies long regulatory elements defined by a specific
region (for example -500/+300 around the TSS) and a specific short
k-mer (for example TAA). For this, \dexter\ computes the correlation
between gene expression and the frequency of the k-mer in the region,
and searches for pairs of (k-mer, regions) that show the highest
correlation with gene expression. All identified pairs are then used
as predictive variables of a linear model trained to predict gene
expression with LASSO penalty (each model involves a few dozens
parameters). The approach was applied to various organisms (malaria
parasite, yeast, worm, human, plant, \dots) in various conditions
(cell types or development stages) and still showed a
surprising accuracy of 60\% correlation between measured and predicted
expression in average. For the malaria parasite, the accuracy even
increases above 70\% in several life-cycle stages, indicating that low
complexity sequences could have a predominant role in the control of
gene expression for this organism. The number of parameters of each
\dexter\ model does not exceed 100 before applying the LASSO penalty.

\subsubsection{Deep learning approaches}
\bigskip
In 2018, Zhou et al. \parencite{zhou_deep_2018} proposed the
\hypertarget{expecto}{\approach{ExPecto}} approach, which capitalizes on the DeepSea method
that the same group previously published (see above section)
\parencite{zhou_predicting_2015}. The approach involves 3 
steps. 1) A CNN model similar to that used in DeepSea is learned on
2,002 genome-wide histone marks, transcription factor binding and
chromatin accessibility profiles (data from ENCODE and Roadmap
Epigenomics projects). 2) These 2,002 models are used to scan 40kb
regions centered on gene TSSs. The score of these CNNs at 200
positions along the 40kb sequences provide a very large set of
$200\times 2,002$ features describing each gene. This set is reduced
to $10\times 2,002$ features by a spatial transformation that uses
exponential decay functions. 3) These features are used as input to
train different linear models that predict the RNA-seq signal
associated with each gene in 218  tissues and cell types
(selected from GTEX \parencite{lonsdale_genotype-tissue_2013}, ENCODE and
Roadmap Epigenomics projects). In each experiment, the accuracy of the
model was measured by the correlation between the predicted and measured
RNA-seq signal. Expecto shows impressive prediction accuracy, with a
median $81.9\%$ correlation across the 218 models. The linear model
involves around 20K trainable parameters, but the improved DeepSea
model used to build the features of the linear model is around 200M
according to our estimate.
\bigskip

Another CNN approach proposed in 2018 was \hypertarget{bassenji}{\approach{Bassenji}}
\parencite{kelley_sequential_2018}. The model takes very large sequences of
131kb in input. The network uses standard convolution/pooling layers,
followed by dilated convolution layers. Contrary to standard
convolutions which realize a weighted sum on the output of adjacent
neurons of the previous layer, in the dilated convolution the sum
involves neurons spaced by several positions. In practice, this allows
modeling motif combinations separated by several bps. Moreover, the
network involves several dilated layers with gaps increasing by a
factor of two, which enables the model to potentially capture
combinations spanning an exponential number of bps. The approach was
applied to various DNase-seq and histone modification ChIP-seq, but
also to 973 FANTOM5 CAGE experiments. For these latter, the goal was
to predict the CAGE signal measured on 128~bp sequences associated
with a TSS. The accuracy was measured by the correlation between
the measured and predicted signal and the authors report an average
correlation of 85$\%$.

\bigskip

Another interesting CNN study was the \hypertarget{xpresso}{\approach{Xpresso}} approach
\parencite{agarwal_predicting_2020}, whose aim was to study the link between
mRNA expression level on one side, and promoter sequence and mRNA
features related to mRNA stability on the other side. The authors
proposed a CNN model that takes in input sequences -7Kb/+3.5Kb
centered on the TSS. The architecture was comprised of two sequential
convolutional and max-pooling layers, followed by two fully connected
layers preceding the output neuron. In addition to the output of the
second convolutional layers, the first fully connected layer takes in
input sequences features commonly associated with mRNA stability,
\ie\ exon density and length, and G/C content of 5'UTR, 3'UTR and
ORF. The authors reported a correlation between measured and predicted
expression signal around 71\% in human K562 and up to 77\% in mouse
mESC. To note, the number of trainable parameters is around 112K
according to the authors.

\bigskip

Several other CNN approaches have been proposed in the last years. For
example, in \parencite{grapotte_discovery_2021}, our group proposed
the \hypertarget{deppstr}{\approach{deepSTR}} model to study the
regulation of the CAGE signal that specifically initiates at
microsatellites\hdronly{ (see Appendix D)}. deepSTR uses 50bp
sequences centered on the 3'end of the microsatellite and shows a
correlation between predicted and measured CAGE signals up to 80\% for
certain classes of microsatellites. Besides pure CNN approaches, one
of the most recent developments in the field was the
\hypertarget{enformer}{\approach{Enformer}} approach
\parencite{avsec_effective_2021} that uses self-attention techniques
developed for natural language processing
\parencite{vaswani_attention_2017}. The Enformer model uses several
standard convolutional layers to identify motifs in the input
sequence. The output of these convolutional layers then goes trough
several \textit{multi-head attention layers} that share information
across the motifs and can model long-range interactions, such as those
between promoters and enhancers. Enformer uses very long input
sequences of 196K~bp and predicts $5,313$ and $1,643$ different
genomic signals (DNAse, ChIP-seq and CAGE expression data) for the
human and mouse genome, respectively. According to the authors, the
total number of trainable parameters of Enformer is around
250M. Compared to Bassenji on several data, Enformer shows moderate
but constant improvement in accuracy.

\section{Identifying and prioritizing genomic variants}\label{s:variants}
Maybe one of the most common applications of machine learning, and
especially deep learning, for regulatory genomics has been the
identification and prioritization of genomic variants from genome-wide
association studies (GWAS) and expression quantitative trait loci
(eQTL) studies. GWAS use statistical methods to identify genetic loci
associated with common diseases and traits. This involves the analysis
of thousands of variants in large cohorts of individuals, often split
into cases and controls, to identify variants associated with the
trait of interest (\ie\ genomes with this variant have often this
trait/pathology). Similarly, eQTL studies identify associations
between genomic variants and the expression of specific genes (genomes
with this variant often show over-expression of gene $X$). GWAS and
eQTL studies have identified thousands of variants
statistically-associated with a physiological trait or an expression
profile, which are available in dedicated databases
\parencite{macarthur_new_2017,aguet_genetic_2017}.  However, because
variants are not independent (an effect called \textit{linkage
  disequilibrium}), many variants are often present together in the
same genomes. As a consequence, it is widely believed that only a few
GWAS or eQTL variants are truly functional
\parencite{aguet_genetic_2017}. In these conditions, an interesting
idea is to use the ML models described above for the identification
and prioritization of functional variants.  In its simplest form, the
procedure involves taking a specific model---predicting for example if
a sequence is bound by a given TF---and computing the prediction of
this model at a given locus using i) the reference genome and ii) the
mutated genome associated with a specific variant (for example a
specific SNP). Variants that increase or decrease significantly the
predicted signal are more likely to have a strong effect on the cell
and hence to be functional
\parencite{alipanahi_predicting_2015,zhou_predicting_2015,quang_danq:_2016,zhou_deep_2018}. Using
this principle, the authors of
\hyperlink{deepbind}{\underline{DeepBind}} proposed a new genomic
representation called \textit{mutation maps} to visualize the effect
that every possible point mutation in a sequence may have on binding
affinity \parencite{alipanahi_predicting_2015}.

More sophisticated approaches have also been proposed to identify
variants using a combination of ML models. For example, a deep
learning model that uses the score of $\sim 600$ DeepBind
transcription factor models has been trained to discriminate between
high-frequency variants (assumed as neutral) and putatively
deleterious variants from GWAS studies
\parencite{alipanahi_predicting_2015}. This model takes in input the score
of the $\sim 600$ DeepBind models both for the wild type sequence
(reference genome) and the mutant sequence. Similarly, a boosted
logistic regression classifier and a boosted ensemble classifier have
been trained to discriminate between high-frequency SNPs and
putatively deleterious variants from GWAS and eQTL studies using the
epigenetic-mark predictions of \hyperlink{deepsea}{\underline{DeepSea}} and \hyperlink{danq}{\underline{DanQ}}, respectively
\parencite{zhou_predicting_2015,quang_danq:_2016}.

\section{Assessing regulatory hypotheses with ML models}\label{s:assessing}
Besides the prioritization of genomic variants, the ML approaches
presented here, along with many others and the concordant development
and availability of various omics data, have enabled several advances
in regulatory genomics over the past decade. As explained in the
introduction, one of the most striking results of these studies is the
fact that gene expression can be predicted with often high accuracy
from the sequence only. Beyond this general result, the supervised
framework offers the possibility to test various hypotheses by simply
training a model and estimating its accuracy on specific
problems. This can be done in various ways. One approach is to
directly control the nature of the input of the model. For example,
with \hyperlink{xpresso}{\underline{Xpresso}}, Agarwal \& Shendure
studied the impact of sequence length on model accuracy
\parencite{agarwal_predicting_2020}. They showed that although the
large promoter sequence -7Kb/+3.5Kb around the TSS provides the best
accuracy, most of the information for the control of mRNA expression
lies in -1.5Kb/+1.5Kb because these sequences provide approximately
the same accuracy as the longer -7Kb/+3.5Kb. Similar results were obtained with the
\hyperlink{deepstr}{\underline{deepSTR}} model for studying the
regulation of the transcriptomic signal that initiates at
microsatellites \parencite{grapotte_discovery_2021}\hdronly{ (Appendix
  D)}. Specifically, the analyses shown that the 50bp sequence centered on
the 3' end of the microsatellites is sufficient to predict the CAGE
signal with often good accuracy. Similarly, by limiting the input
features to the frequency of short kmers on specific long DNA regions,
we have shown with \hyperlink{dexter}{\underline{\dexter}} that low
complexity regions may play an unsuspected role in gene expression
regulation of \Pf\ and several other eukaryotes
\parencite{menichelli_identification_2021}\hdronly{ (Appendix C)}.

Another procedure to test different hypotheses is to control the way
positive and negative examples are selected. For example, in
\parencite{zheng_deep_2021} the authors study TF binding with a
CNN. The originality of this study is that the sequences are selected
in a way that insures that the TF motif is present both in the
positive (bound) and negative (unbound) sequences.  In this way, the
predictive model cannot simply use the presence/absence of the target
motif to discriminate the sequences, and the good accuracy of the
approach showed that other features are likely involved in the binding
site recognition mechanism of the studied TFs. To go one step further,
the authors restricted all sequences to be within DNaseI
hypersensitive sites (a mark of accessible chromatin). They trained a
new model and observed a drop in the accuracy, suggesting that DNA
features related to open chromatin are likely involved in this
recognition process.

Rather than controlling the sequences used for learning the model,
another simple technique that can be used to test some hypotheses is
to compare the accuracy achieved by a given model on different sets of
sequences. For example, Agarwal \& Shendure studied the impact of
enhancers on the prediction of \hyperlink{xpresso}{\underline{Xpresso}}
\parencite{agarwal_predicting_2020}. Xpresso only uses the sequence located
around the TSS for the prediction and hence cannot handle distal
regulatory elements. Given that enhancers are frequently associated
with large domains of H3K27Ac activity, the authors identified a set
of 4,277 genes that overlap H3K27Ac marks and observed that the
predicted expression level on these genes is markedly smaller than the
measured expression, a trend that was not observed for the other
genes. Hence, the expression of genes under enhancer control seems to
be underestimated by the Xpresso model, which provides a practical
way for identifying these genes.

Finally, a very useful tool available in the ML toolbox is swap
experiments. Model swapping is used to test hypotheses related to the
specificity of a regulation mechanism. The procedure involves learning
different models on different conditions (says conditions A and B),
and to compare the accuracy of the model learned on A when it is
applied to B, to the accuracy of the model learned and applied on
B. In practice, an important point to which practitioners must pay
close attention is that the sequences used for testing in one
condition were not used for training in another one, to avoid any bias
in the accuracy estimates during swaps. This technique was extensively
used in the \hyperlink{tfcoop}{\underline{TFcoop}} analyses
\parencite{vandel_probing_2019}\hdronly{ (Appendix B)}. Given two
ChIP-seq experiments targeting the same TF in two cell types, a TFcoop
model predicting the binding was learned for each cell type. Then the
two models were swapped. In some cases, both models got the same
accuracy, but for some TFs the model learned on cell type A was not as
good on cell type B as the model directly learned on cell type B,
meaning that the co-factors of the TF were likely not the same in the
two cell types. Similarly, for a given ChIP-seq experiment, a model
was learned for enhancer sequences and another model for promoter
sequences. The swap experiment revealed that the two models are not
interchangeable, and hence that the co-factors of a TF may vary
depending on the nature of the regulatory sequence (promoters
\vs\ enhancers). On the contrary, models for mRNA promoters, lncRNA
promoters and pre-miRNA promoters appeared to be perfectly
interchangeable. Agarwal \& Shendure also did swap experiments between
human and mouse \parencite{agarwal_predicting_2020}. Namely, they
trained one \hyperlink{xpresso}{\underline{Xpresso}} model on mouse
and one on human using comparable RNA-seq experiments, and observed on
a set of orthologous genes that the model learned on mouse has the
same accuracy on human as the model directly learned on human (and
conversely), meaning that the regulatory principles are likely very
close in both organisms.

\section{Breaking the regulatory code: Interpreting ML models}\label{s:interpreting}

Finally, the most ambitious application of ML is obviously to break
the regulatory code and determine the rules used by the cell to
regulate expression. A first step toward this goal is to identify all
regulatory elements involved in the regulation process in
question. Note that the term \textit{regulatory element} has to be
understood in its broadest sense here.  It may stand for standard
motifs of TF binding sites, but it can also refer to low complexity
regions such as CpG islands \parencite{deaton_cpg_2011} or short
tandem repeat \parencite{gymrek_abundant_2016} that are also involved
in gene regulation, as well as any other kind of ``sequence patterns''
that we can think of, such as, for example, the DNA shapes proposed by
\parencite{rohs_role_2009}. In addition to the regulatory elements,
breaking the regulatory code also means determining how these elements
are combined, and what rules they follow in terms of repetition,
position, and orientation on the DNA sequence.  Getting down to this
level of detail involves analyzing the learned model to understand how
it works. This is referred to as \textit{model interpretation} in the
ML community. There are several definitions of interpretability in
the ML literature, and these definitions are often considered as
domain-specific
\parencite{doshi-velez_towards_2017,lipton_mythos_2018,gilpin_explaining_2019,rudin_stop_2019}.
In the context of regulatory genomics, model interpretation has also
several meanings, and we will see that the term covers different
approaches that do not provide the same kind of information. The
general problem is as follows: we have a predictive model that has
been learned for a specific problem, and we want to understand how it
works. There are at least three different ways to attempt this. The
first and more direct approach is to explain the model by analyzing
its different components and extracting the rules used to make its
predictions. A second approach is to explain the predictions of the
model rather than the model itself, \ie\ given a sequence and a
prediction, we want to explain why the model predicted this value for
this sequence.  A third alternative is to try to explain the model by
exploring its behavior on synthetic sequences. Theoretically, the
three approaches are possible for any kind of model. However, as we
will see, linear models and random forests can often be partly
explained by a direct analysis of their components.  On the contrary,
this is more difficult to do for models based on CNNs, so different
methods belonging to the second and third categories have been
developed for these laters.

\subsection{Explaining model components}
In some cases, the model may be simple enough to be directly analysed
as a whole. This view of interpretability argues for simple and sparse
models (\ie\ with few parameters). This may be possible for some very
small decision trees and linear models, but this is generally not the
case for the regulatory models described above. Moreover, even for a
simple linear model, it is known that interpretation is not obvious, as
the sign of the coefficient associated with a specific variable (which
is tempting to interpret as the sign  of the association between this
variable and the predicted signal) may  change depending on
the identity of the other variables included in the model.

A less stringent view of interpretability corresponds to models that
can be broken down into different modules that are further analyzed
\parencite{lipton_mythos_2018,gilpin_explaining_2019}. For regulatory
genomics, this corresponds to the case where we can extract the DNA
features used by the model for its predictions. Note that a DNA
feature may be more complex than simply the identity of a regulatory
element.  For example, in the \hyperlink{dexter}{\underline{DExTER}} method, the predictive features
also encode information about the position of the regulatory elements
on the DNA sequence
\parencite{menichelli_identification_2021}\hdronly{ (Appendix
  C)}. Hence, DNA features may potentially encode complex information
related to the number of repetitions, position, and orientation of
regulatory elements on the DNA sequence.  Obviously, extracting the
DNA features used by a model is relatively easy to do for all models
that directly take these DNA features in input, such as
logistic/linear models and random forests. On the contrary, for CNN
approaches the task is more difficult as we will see below.

\subsubsection{Random forests and linear models}
In these models, extracting the DNA features involves identifying
which variables provided in the input are the most important for the
predictions. \hyperlink{expecto}{\underline{ExPecto}} is an example of linear model where such
interpretation is quite easy \parencite{zhou_deep_2018}. In this
paper, the authors observed for example that, in most cases, the
ExPecto models predict expression with features related to
transcription factors and histone marks, but not with DNase I sequence
features. Note that the DNA features extraction may be further
facilitated if the learning algorithm directly includes a feature
selection procedure such as the LASSO
\parencite{tibshirani_regression_1994,zhao_model_2006}, which
drastically decreases the number of predictive features used by the
model. \hdronly{This is the approach we followed in several of our studies
about transcriptomic regulation \parencite{bessiere_probing_2018,
  menichelli_identification_2021} (Appendix A, C) and TF
binding
\parencite{vandel_probing_2019,romero_systematic_2022}
  (Appendices B, E)}.

Rather than simply providing the list of features used by the model,
it is much more useful to also provide a measure of importance
associated with each feature
\parencite{murdoch_definitions_2019,williamson_general_2021,benard_mean_2022}. For
random forests, this is usually done by shuffling the values of each
variable. Namely, the accuracy of the model is first estimated on the
test set. Then, the values of the $i$th variable are randomly
shuffled, and the loss of accuracy induced by this noise is used as a
measure of the importance of the $i$th variable in the model
\parencite{breiman_random_2001,benard_mean_2022}. By repeating this
process, the relative importance of each variable is estimated. A very
similar procedure also exists for linear/logistic models. In this
procedure, the value of the coefficient associated with the $i$th
variable is set to zero (which is equivalent to simply removing this
variable from the model) and, as for random forests, the loss of
accuracy induced by this operation is used as a measure of variable
importance.  In both cases, this measure can be considered as
objective, as it is based on the loss of accuracy when removing
information carried by each variable. This approach was extensively
used in \hyperlink{tfscope}{\underline{TFscope}} for analyzing the
binding differences of one TF in two conditions (two cell types or two
treatments), or of two paralogous TFs with similar PWMs
\parencite{romero_systematic_2022}\hdronly{ (Appendix E)}. For this,
the TFscope model integrates three kinds of variables that model i)
the core motif, ii) the nature and position of binding sites of
co-factors, and iii) the nucleotidic environment of specific regions
around the core motif. The importance of each feature in the different
comparisons was assessed with the importance measure described
above. For comparisons involving one TF in two conditions, the
co-factors, and to a lesser extent the nucleotidic environment, were
often (but not always) the most important features explaining the
differences of binding sites. For paralogous TFs (two TFs in the same
cell type) the picture is different, and subtle differences in the
core motif often explained most binding-site differences.

\subsubsection{CNNs}
Approaches based on CNNs take raw sequences in input, so the DNA
features are not directly provided to the model. However, as we have
seen, the filters of the first convolutional layer  correspond
to PWMs modeling DNA motifs. Hence, we could think that the DNA
features used by the model are actually encoded in the first
convolutional layer and that we could extract these filters to get
the information. It is however not so simple in practice.  Koo \& Eddy
studied the way CNNs build representations of regulatory genomic
sequences \parencite{koo_representation_2019}. They showed that the filters
of the first layer actually encode partial motifs which are then
assembled into whole DNA features in the deeper layers of the
CNN. Hence, extracting the DNA feature associated with a specific
filter is not as immediate. However, some attempts have been made in
this direction. The authors of \hyperlink{deepbind}{\underline{DeepBind}} proposed an interesting
approach where all sequences that pass the activation threshold of the
filter of interest are extracted and aligned on the position producing
the maximum activation signal for this filter
\parencite{alipanahi_predicting_2015}. Then a PWM of predefined length $m$
is learned from this alignment using standard PWM-learning methods
\parencite{wasserman_applied_2004}. The same approach was also used for the
\hyperlink{danq}{\underline{DanQ}} model \parencite{quang_danq:_2016}. The reconstructed PWMs were then
aligned to motifs available in known-motif databases
\parencite{fornes_jaspar_2020} using the TOMTOM algorithm
\parencite{gupta_quantifying_2007}. Of the 320 filters learned by the DanQ
model, 166 significantly matched known motifs. Hence, extracting the
PWMs learned by a CNN seems possible at least partly, although the
above approach does not completely warrant against the ``partial
motif'' problem. An open question that remains is how to associate a
measure of importance to these PWMs? Theoretically, a CNN filter
can be easily turned off, so it could be possible to estimate the
accuracy of the CNN with and without the filter to have a measure of
the importance of this filter for the predictions. However, as the CNN
may model the same motif using different filters (or even using
several filters that are combined in the deeper layers), turning a
filter off does not warrant that the associated motif is also off,
which may lead to underestimating its importance.

\subsection{Explaining model predictions}
Because of the partial motif problem, directly explaining a CNN
remains a difficult question. Hence several methods have been proposed
to explain the predictions of the model rather than directly
explaining the model. This involves using the model on a specific
sequence and identifying
the nucleotides that have the highest weight for
the computation of the predicted value. The simplest method (known as
\emph{input perturbation} or \emph{\textit{in silico} mutagenesis}) is
similar to the mutation maps proposed for variant identifications. It
involves systematically simulating every single-nucleotide
perturbation of the input sequence, and recording the effect induced
on the predicted value
\parencite{alipanahi_predicting_2015,zhou_predicting_2015,zheng_deep_2021}.
Computing model prediction on every variant that can be obtained from
single-nucleotide perturbations of a sequence may however induce high
computational cost, specifically if the sequence is long, or if the
analysis is repeated over many sequences. Backpropagation-based
approaches have been proposed as more computationally efficient
alternatives. The idea is to propagate an importance signal from the
output neurons to the inputs through the different layers of the model
using a backpropagation algorithm. In this way, a single pass is
sufficient to compute the contribution of each nucleotide for the
computation of the output value (these contributions can then
visualize using \emph{saliency maps}). Several approaches have been
proposed on this idea
(\eg\ \parencite{simonyan_deep_2014,lanchantin_deep_2017,shrikumar_learning_2019}). In
\parencite{zheng_deep_2021}, a simulation-based study is used to benchmark
these different approaches in the context of regulatory genomics.

By looking at the contribution scores of each nucleotide, one can
identify the sub-sequences and regions that appear to be the most
important for the prediction. Moreover, these sub-sequences can be
further compared to known motifs to identify potential TFs involved in
the regulation of the predicted signal. Hence, these approaches may
highlight potentially interesting genomic information hidden into a
specific sequence.  However, it is important to note that because
contribution scores are computed independently for each nucleotide,
there is no warranty that the most important regions identified this
way are really the most important regions for CNN prediction.
Moreover, these analyses are restricted to some specific sequences and
are not sufficient to understand the general mechanisms used by the
cell to regulate gene expression at different loci.  For this, further
analyses are needed. A possible approach is to compute the
contribution of every nucleotide of all positive sequences and to
extract the recurrent DNA motifs from these contribution maps. A first
method is to use these maps to compute the average importance of every
k-mer of a fixed length. The most important k-mers are then identified
and compared to known TF motifs using approaches like TOMTOM
\parencite{gupta_quantifying_2007}. This is for example the approach
used in \parencite{zheng_deep_2021} to identify several co-factors
that likely explain the binding differences between positive and
negative sequences of 38 TFs. Another more sophisticated approach is
to directly learn DNA motifs (using PWMs or close models) from the
contribution maps. This is the aim of the TF-MoDISco method that
segments the contribution maps into seqlets (a kind of weighted
k-mers), and then clusters the seqlets in different motifs
\parencite{shrikumar_technical_2020}. This approach has been used in
\parencite{avsec_base-resolution_2021} to identify core motifs and
potential co-factors in four ChIP-seq experiments.

Thus, different approaches based on prediction explanation can also be
used to identify motifs that are likely encoded into the structure of
the CNN. However, in our opinion an important difference with
approaches based on the analysis of model components is that these
methods lack an objective measure of importance associated with each
motif. Indeed, it is not possible to turn the discovered motifs off in
the CNN, and thus we cannot measure the accuracy of the model with and
without a motif. Another problem is that there is no warranty that all important
motifs used by the CNN were identified by the extraction
procedure. Actually, it is even possible that some regulatory elements
learned by the CNN but which were not identified by the motif
extraction approach---for example because these elements do not fit
the kind of motifs searched by the extraction procedure---are actually
more important for model predictions than the identified motifs.

\subsection{Exploring model behavior on synthetic sequences}
In the above sections, we have seen methods to extract some of the motifs
that have been learned by a CNN for predicting a given
signal. However, these interpretation methods are limited to the
identification of simple motifs and cannot handle more sophisticated
DNA features, such as the number of repetitions of a motif, the
combination of several motifs, or their relative position on the sequence. For
this, other approaches based on synthetic sequences have been
proposed.

The first approach for this was the DeMo dashboard
\parencite{lanchantin_deep_2017}, which was inspired by the work of
\parencite{simonyan_deep_2014} to interpret CNNs in the context of image
recognition. DeMo aimed to predict TF binding with a CNN. To
identify the genomic features captured by the CNN, the authors
proposed to construct the best synthetic sequence that would maximize
the binding probability according to the CNN. The idea was to provide
the users with an archetype of the sequences with the highest binding
probability, which supposedly bears the DNA features required for the
binding. The idea was interesting but may miss certain features,
especially if several (exclusive) rules govern the binding.

To go one step forward, \parencite{koo_global_2021} proposed an approach
named \textit{global importance analysis} that samples a large number
of synthetic sequences with and without a specific DNA feature, and
uses the learned CNN as an oracle to predict the binding associated
with each sequence. If the predicted binding signal is statistically higher for
the sequences that embed the studied DNA feature than for the others,
the feature is considered important. The approach is interesting in
that it can be used to assess any feature. For example,
\parencite{koo_global_2021} used it to show that increasing the number of
repeats of the RBFOX1 motif increases the binding probability of the sequence
according to their model. Similarly, they also showed that including some GC bias
in sequence 3'end also increases this probability. The approach was also used
in \parencite{avsec_base-resolution_2021} to study the sinusoidal pattern
that seems to regulate the distribution of the binding sites of the
Nanog TF and its co-factors along the DNA sequence. It is important to
understand that this approach is not a fully automatic method that
would extract all DNA features captured by a CNN. Rather, specific
hypotheses have to be constructed before being assessed by generating
appropriate synthetic sequences. In this sense, the approach is
similar to that of the ML models that directly take meaningful DNA
features in input: in both cases, one can discover solely what has been
formerly hypothesized.

One advantage of this approach is that it can be used to estimate the
relative importance of different features. Note however that this
measure of importance is somewhat subjective, as it is only dictated
by the model, with no link with its accuracy. One issue is that the
approach does not comply with a fundamental assumptions in ML, which
states that the training and test sets must be representative of the
samples to classify.  Here, the synthetic sequences do not follow
exactly the same distribution as the sequences used to train the model
and to estimate its accuracy. As a result, this accuracy may be an
optimistic estimate of the accuracy of the model on the synthetic
data. Say differently: even if the model is quite good on the test
sequences, it may be inaccurate on the synthetic sequences. This is
related to the notions of extrapolation \vs\ interpolation. While ML
models can be good at predicting a signal associated with a sequence
that is close to other sequences in the training set (interpolation
problem), when the sequence is far from any example of the training
set (as the synthetic sequences can be) the problem is obviously more
difficult, even if CNNs seem to be able to handle a certain level of
extrapolation \parencite{balestriero_learning_2021}. Note that the
approach can nonetheless propose interesting hypotheses, which then
need to be validated experimentally; however, users/experimentalists
should be aware that the probability that these hypotheses holds can be
lower than the model accuracy may lead one to believe.

\section{Conclusions}
Over the past decade, several ML models have been applied to the
modeling of long regulatory sequences and have led to substantial
advances in our understanding of regulatory genomics. These studies
take place in a supervised framework, which enables, among other
advantages, a fair comparison of model performance in terms of
accuracy. From this perspective, deep learning approaches based on
CNNs often show substantial improvement over simpler methods. This is
especially true for the prediction of gene expression signal, where
many efforts have been made to allow CNNs to handle very long
sequences to capture the effect of potential distant enhancers.

Besides model accuracy, we have seen that different level of knowledge
can be gained by these studies.  The prioritization of nucleotide
variants, which is of great interest in medical and therapeutic studies,
can be deduced from model predictions without the need for sophisticated procedures
of model interpretation. Similarly, general hypotheses about the
specificity or the extent of a regulatory mechanism can often be
tested by training and swapping different models, or by changing the
training or the test sets. On the contrary, identifying complex DNA
features requires knowledge-extraction procedures that must be
specifically adapted to the model at hand. The task is obviously
easier for ML models that directly take these features in input than
for CNN-based models that learn the predictive features from raw
sequences. However, it would be inaccurate to consider these latter as
complete black boxes, as several approaches are now available to
extract such information. Notably, the oracle approach based on
synthetic sequences looks like a promising avenue
\parencite{koo_global_2021}, provided that the generated sequences
remain close to the training sequences to avoid misleading
conclusions. Actually, in our opinion, the main limitation of CNN
models lies in the difficulty to globally and objectively assess the
importance of the extracted features. A related issue is the fact that
we do not know with precision whether all important features have been
extracted or if the CNN actually uses additional important features
that were not identified \parencite{murdoch_definitions_2019}.

We have presented in this \doc\ some approaches for
identifying the DNA features used by a specific model to make its
predictions. A question that is rarely addressed in these studies is
the stability of the model. DNA features are often highly
correlated. For example, motifs of TFs of the same family are often
very close. Moreover, motifs of co-factors often appear together at
the same loci. Similarly, certain motifs are strongly enriched in DNA
sequences with specific nucleotide or dinucleotide content. These
strong correlations induce that different models with different DNA
features may have close accuracy. As a result, the model that has
been learned is not necessarily the only ``good'' model, a problem
sometimes referred to as the \textit{Rashomon effect} in ML (after the
1950 japanese movie directed by Akira Kurosawa)
\parencite{breiman_statistical_2001,fisher_all_2019}. From this
perspective, by aggregating the predictions of many models, ensemble
ML approaches---and hence random forests---are less prone to this
issue \parencite{breiman_statistical_2001}. For other methods,
analysis of model stability can help
\parencite{meinshausen_stability_2010,piles_feature_2021,nogueira_stability_2018}. However,
this requires repeating the learning procedure several times using
slightly different learning sets, which can involve prohibitive
computing time for very complex models.

\section*{Acknowledgements}
I thank Charles Lecellier, Sophie Lèbre and Olivier Gascuel for their
careful reading of this manuscript. I am also grateful to Charles and
Sophie for our long-standing collaboration in the field. Some of the
views developed here undoubtedly stem from our daily discussions.

\section*{Funding}
This work was supported by the SIRIC and the Labex NUMEV (projet
MOTION), the National Research Agency (ANR project LOWCO) and by the
LabMUSE EpiGenMed (R-loops project).

\bibliography{HDR}
\end{document}